\documentclass[runningheads]{llncs}
\usepackage[utf8]{inputenc} 
\usepackage[T1]{fontenc}    
\usepackage{hyperref}       
\usepackage{booktabs}       
\usepackage{amsfonts}       
\usepackage{nicefrac}       
\usepackage{microtype}      
\usepackage[pdftex]{graphicx}
\usepackage{pgfplots}
\usepackage{pgfplotstable}
\pgfplotsset{compat=1.7}
\usepackage{tikz}
\begin{document}
\title{SerumRNN: Step by Step Audio VST\\ Effect Programming}


\author{Christopher Mitcheltree\inst{1, 2}\orcidID{0000-0002-2844-3650} \\ \and
Hideki Koike\inst{1}\orcidID{0000-0002-8989-6434}}

\authorrunning{C. Mitcheltree and H. Koike}

\institute{Tokyo Institute of Technology, Tokyo 152-8550, Japan\\ 
\email{christhetree@gmail.com}\\
\email{koike@c.titech.ac.jp}\\
\url{https://www.vogue.cs.titech.ac.jp/} \and
Qosmo Inc, Tokyo 153-0051, Japan\\
\url{https://qosmo.jp/}}

\maketitle              

\begin{abstract}
Learning to program an audio production VST synthesizer is a time consuming process, usually obtained through inefficient trial and error and only mastered after years of experience. As an educational and creative tool for sound designers, we propose \emph{SerumRNN}: a system that provides step-by-step instructions for applying audio effects to change a user's input audio towards a desired sound. We apply our system to Xfer Records Serum: currently one of the most popular and complex VST synthesizers used by the audio production community. Our results indicate that \emph{SerumRNN} is consistently able to provide useful feedback for a variety of different audio effects and synthesizer presets. We demonstrate the benefits of using an iterative system and show that \emph{SerumRNN} learns to prioritize effects and can discover more efficient effect order sequences than a variety of baselines. 

\end{abstract}

\section{Introduction and Background}

Sound design is the process of using a synthesizer and audio effects to create a desired output sound, typically by leveraging virtual studio technology (VST) on a computer. Often, the audio effects applied to the synthesizer play the biggest role in producing a desired sound. Sound design for the music industry is a very difficult task typically done by professionals with years of experience. Educational tools are limited and beginners are usually forced to learn via trial and error or from online resources created by others who typically also learned in a similar way. This makes the learning curve for sound design very steep.

\subsection{Serum}

Serum is a powerful VST synthesizer made by Xfer Records \cite{ref_serum} that can apply up to 10 audio effects to the audio it generates. Serum is currently one of the most popular VST synthesizers in the audio production community and is routinely used by hobbyists and professionals alike. We chose Serum because we wanted to apply our research to a relevant, widely adopted, fully-featured synthesizer that is challenging for humans to master and will therefore maximize the practicality of our work. 

\subsection{Related Work}

While applying AI to sound design is a relatively niche research area, there has been some prior work on leveraging AI to program audio VSTs. K-means clustering + tree-search \cite{ref_k_means} and evolutionary algorithms such as genetic algorithms \cite{ref_ga_1,ref_ga_2_neural_1} and genetic programming \cite{ref_ga_3_gp_1} have been applied to this problem with varying levels of success. Genetic algorithms have also been used to model audio effects directly \cite{ref_genetic_reverb}. However, these systems suffer from one or more of the following problems: 

\begin{itemize}
\item They are applied to toy VSTs with little practical use.
\item They are incompatible with existing VST plugins.
\item Their inference time is prohibitively long.
\item They are black-boxes with uninterpretable results.
\end{itemize}

With the recent rise in deep learning and neural networks, there has also been some related work using deep convolutional neural networks (CNNs) to program audio VSTs \cite{ref_inversynth,ref_ga_2_neural_1}. \emph{InverSynth} \cite{ref_inversynth} is probably most similar to \emph{SerumRNN} since it also makes use of CNNs to program synthesizer parameters. However, as mentioned previously, these neural systems approach the problem with a one-shot, black-box process and focus more on synthesis rather than applying effects. This end-to-end approach replaces a user more than it augments them and results in fewer opportunities to learn. Most users typically know how to begin programming their desired sound via oscillator, attack, decay, sustain, and release parameters, but have difficulty programming the relevant effects due to the sheer number of combinatorial possibilities. 

Finally, there has also been research on using deep learning to model audio effects and / or applying it directly to raw audio \cite{ref_direct_2,ref_ddsp,ref_direct_1,ref_direct_3}. These systems typically use interesting signal processing techniques and neural network architectures from which we draw inspiration. However, they typically cannot be applied to existing VST synthesizers making their usefulness for our goals limited.

Overall, we believe that when using an AI assisted system, the user's sense of ownership over their work should be preserved. As a result, \emph{SerumRNN} is inspired by step-by-step white-box automatic image post-processing systems \cite{ref_exposure} and collaborative production tools \cite{ref_performance,ref_neural_loops} that can educate and augment a user rather than aiming to replace them.

\subsection{Contributions}

\begin{figure}[h]
\centering
\includegraphics[width=1.0\linewidth]{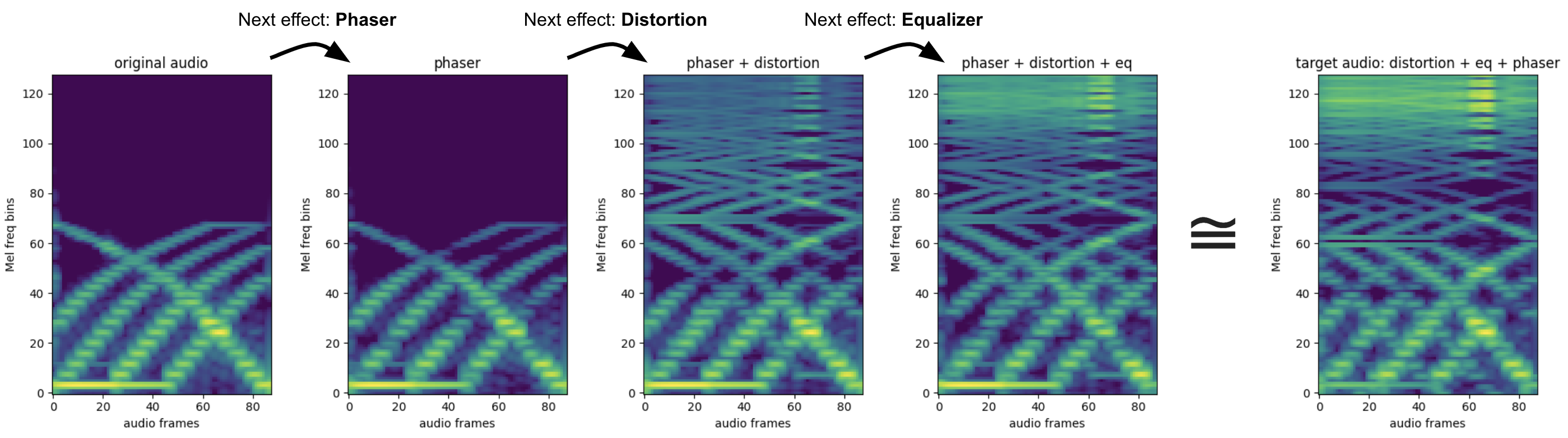}
\caption{Mel spectrogram progression of our system applying three effects to a user's input audio.}
\label{fig_progress}
\end{figure}

In this paper we propose a system (\emph{SerumRNN}) that iteratively changes an input audio towards the same timbre of a desired target audio by sequentially applying audio effects via the Serum VST synthesizer.  As a result, the system provides a sequence of interpretable intermediate steps. It uses, to the best of our knowledge, a novel approach consisting of an ensemble of models working together: an \emph{effect selection model} to determine which effect to apply next to the input audio and then a collection of \emph{effect parameter models}, one per supported effect, to program the selected effect parameters. We demonstrate through extensive evaluation that \emph{SerumRNN}:

\begin{itemize}
  \item Significantly reduces the error between the input and target audio.
  \item Benefits from applying effects iteratively in a specific order.
  \item Learns which effects are most important.
  \item Provides interpretable and valuable intermediate steps.
  \item Can discover more efficient effect order sequences than a variety of baselines.
\end{itemize}

An example of \emph{SerumRNN} applying three steps to some input audio can be seen in Figure \ref{fig_progress}. Audio examples can be listened to at \url{https://bit.ly/serum_rnn}.

\section{Data Collection}

Data collection and processing systems represent a significant portion of the software engineering required for our system. Training data for all models is generated by rendering audio samples from Serum and then converting them into spectrograms and cepstra. Due to the complexity of Serum and its effects, virtually infinite amounts of training data can be generated.

\subsection{Audio Rendering}

\begin{table}[h]
  \caption{Parameters sampled from the Serum VST synthesizer.}\label{tab_params}
  \label{tab_params}
  \centering
  \begin{tabular}{llll}
    \toprule
    Effect & Parameter Name & Type & Sampled Values \\
    \midrule
    Compressor & Low-band Compression & Continuous & [0.0, 1.0] \\
    Compressor & Mid-band Compression & Continuous & [0.0, 1.0] \\
    Compressor & High-band Compression & Continuous & [0.0, 1.0] \\
    Distortion & Mode & Categorical & 12 classes \\
    Distortion & Drive & Continuous & [0.3, 1.0] \\
    Equalizer & High Frequency Cutoff & Continuous & [0.50, 0.95] \\
    Equalizer & High Frequency Resonance & Continuous & [0.0, 1.0] \\
    Equalizer & High Frequency Gain & Continuous & [0.0, 0.4] and [0.6, 1.0] \\
    Phaser & LFO Depth & Continuous & [0.0, 1.0] \\
    Phaser & Frequency & Continuous & [0.0, 1.0] \\
    Phaser & Feedback & Continuous & [0.0, 1.0] \\
    Hall Reverb & Mix & Continuous &  [0.3, 0.7] \\
    Hall Reverb & Low Frequency Cutoff & Continuous &  [0.0, 1.0] \\
    Hall Reverb & High Frequency Cutoff & Continuous &  [0.0, 1.0] \\
    \bottomrule
  \end{tabular}
\end{table}

Five commonly used and predominantly timbre altering effects are chosen for training data collection: multi-band compression, distortion, equalizer (EQ), phaser, and hall reverb.  Table \ref{tab_params} summarizes which Serum synthesizer parameters are sampled for each supported effect. Continuous parameters (knobs on the Serum synthesizer) can be represented as floating-point numbers between zero and one inclusively. Categorical parameters can be represented as one-hot vectors of length \(C\) where \(C\) is the number of classes. Continuous parameter sampling value ranges are occasionally limited to lie within practical, everyday use regions.

Furthermore, in order to apply the system to a variety of different sounds, we collect data from 12 different synthesizer presets split into three groups (in increasing order of complexity): \emph{Basic Shapes}, \emph{Advanced Shapes}, and \emph{Advanced Modulating Shapes}. The \emph{Basic Shapes} preset group consists of the single oscillator sine, triangle, saw, and square wave default Serum presets. Next, the \emph{Advanced Shapes} preset group consists of the dry (no effects) dual oscillator \texttt{"LD Power 5ths"}, \texttt{"SY Mtron Saw"}, \texttt{"SY Shot Dirt Stab"}, and \texttt{"SY Vintage Bells"} default Serum presets. Finally, the \emph{Advanced Modulating Shapes} preset group consists of the dry dual oscillator \texttt{"LD Iheardulike5ths"}, \texttt{"LD Postmodern Talking"}, \texttt{"SQ Busy Lines"}, and \texttt{"SY Runtheharm"} default Serum presets. These final four presets also use intense modulations on top of their use of dual oscillators.

Since five different effects are supported by the system, there are 32 different combinations possible when applying a minimum of zero and a maximum of five effects to an input audio signal. For each of these combinations, a modified automated VST rendering tool \cite{ref_renderman} is used to render and save 4000 mono audio clips. Parameters are sampled randomly and uniquely from the permissible values shown in Table \ref{tab_params} and effect order when multiple effects are used is randomized. All audio samples are played and rendered for one second using a MIDI pitch of C4, maximum note velocity, and a sampling rate of 44100 Hz. This process is repeated for each of the 12 presets resulting in 128k audio clips per preset, 384k audio clips per preset group, and 1.536M audio clips total. We only sample a single C4 pitch due to the inclusion of high quality timbre-preserving pitch shifting algorithms in all commonly used digital audio workstations, thus allowing users to easily warp a desired sound they would like to use as input to the system to C4. This preprocessing step could also be done by the system automatically.

\subsection{Audio Processing}

Audio clips are converted to Mel spectrograms using a Short-time Fourier transform (STFT) with a hop length of 512 samples, a fast Fourier transform (FFT) window length of 2048, a minimum and maximum frequency of 20 Hz and 16 kHz respectively, and a Mel filter-bank of size 128. For each spectrogram the first 30 Mel-frequency cepstral coefficients (MFCCs), representing the smoothed spectral envelope, are also computed and saved. Lastly, each spectrogram is converted to decibels (dB). We use both spectrograms and cepstra in order to provide the subsequent neural networks as much volume, pitch, timbre, and temporal information as possible. Early prototypes of the system using only spectral or cepstral features benefited from including both.

\begin{figure}[h]
\centering
\includegraphics[width=1.0\linewidth]{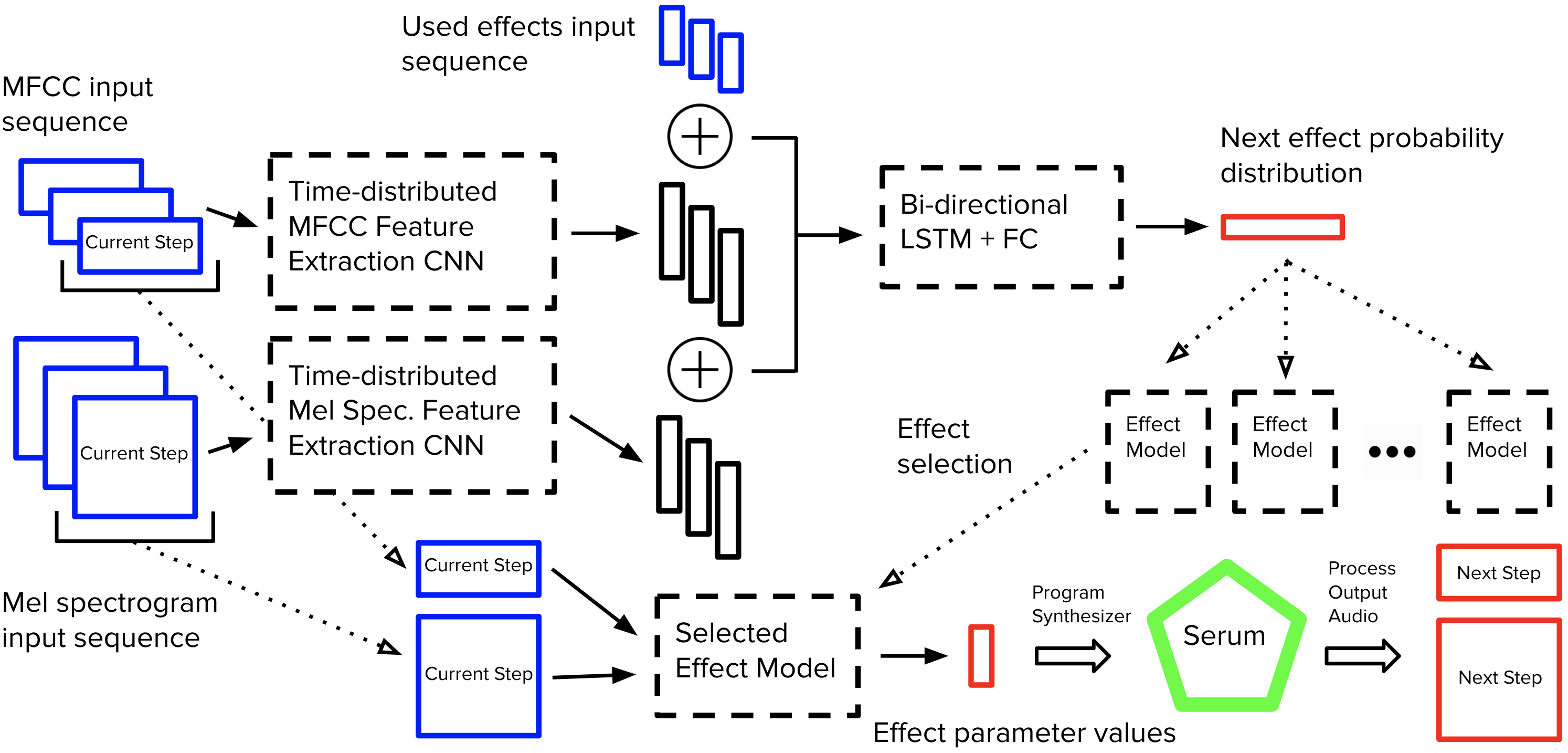}
\caption{System diagram: model components are dashed rectangles, inputs are blue, outputs are red, and the Serum synthesizer is green. Dotted arrows represent the conceptual flow of the system and solid arrows represent the flow of tensors.}
\label{fig_system}
\end{figure}

\section{Modeling}

Our system consists of an ensemble of models that work together to take iterative steps to transform some input audio towards the timbre of some target audio. First, an \emph{effect selection model} determines which effect to apply next and then a collection of five \emph{effect parameter models}, one per supported effect, program the synthesizer accordingly. Figure \ref{fig_system} depicts a diagram of the entire system performing one step.

\subsection{Effect Parameter Models}
\label{sect_effect_param_modeling}

\begin{figure}[h]
\centering
\includegraphics[width=1.0\linewidth]{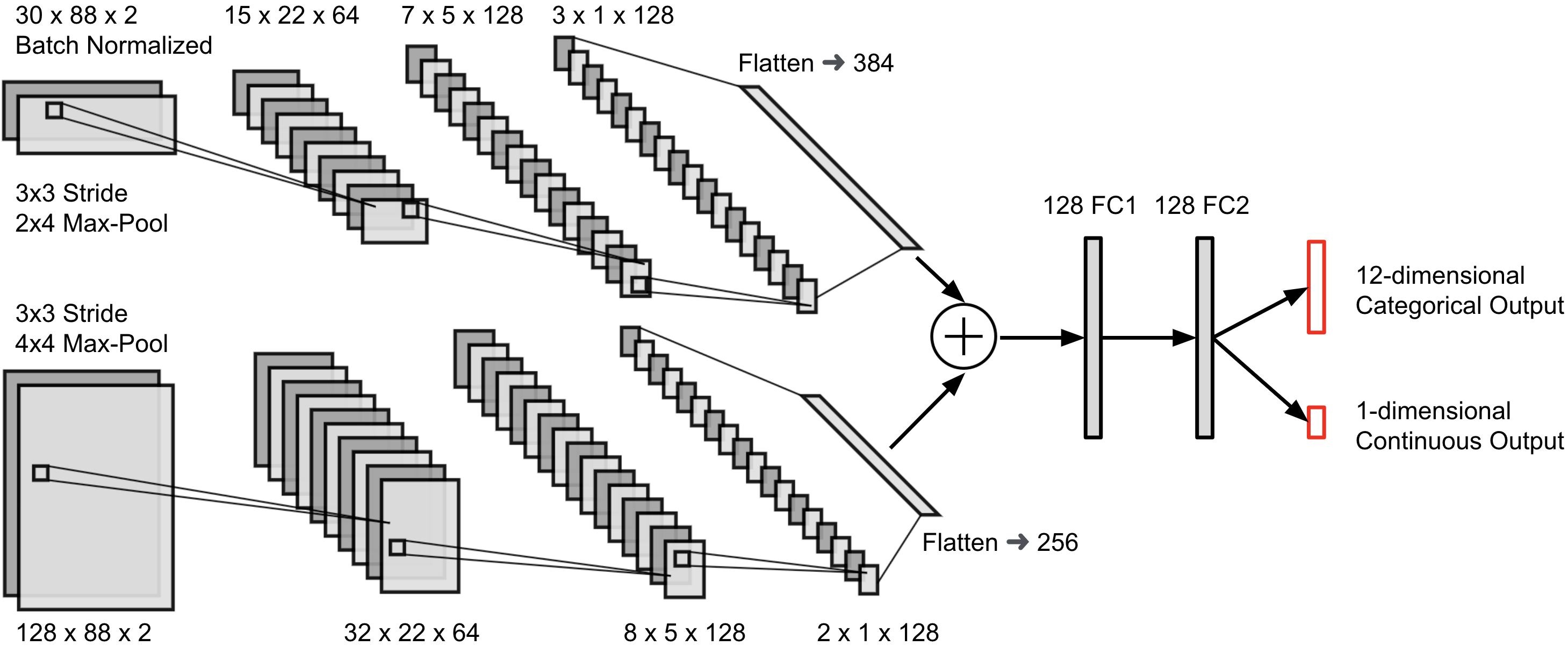}
\caption{\emph{Distortion} effect parameter model architecture (not to scale).}
\label{fig_distortion_arch}
\end{figure}

The output values of effect parameter models are used to program the parameters of an effect. They take two tensors as input: a Mel spectrogram and the corresponding first 30 MFCCs. Both of these tensors consist of two channels, the first representing the input audio and the second representing the target audio. The MFCC tensor is normalized by feeding it through a batch normalization layer that has learnable shift and scale parameters. Two almost identical convolutional neural networks (CNNs) are used to extract features from each input. The Mel spectrogram CNN consists of three convolutional layers each with a 3x3 stride, 4x4 max-pool layer, and 64, 128, and 128 filters respectively. The MFCCs CNN is identical except, due to the smaller \(X\) input dimension, it uses 2x4 max-pool layers. The final max-pool layers of the two CNNs are flattened and concatenated together. This is then followed by two 128-dimensional fully connected (FC) layers with a dropout rate \cite{ref_dropout} of 50\% each. All layers use ELU activations \cite{ref_elu}.

The output layers and loss functions of an \emph{effect parameter model} vary depending on what effect is being modeled. Continuous parameters of an effect, where \(N\) is the number of continuous parameters, are represented as a single \(N\)-dimensional fully connected layer with a linear activation. Categorical parameters of an effect, where \(C\) is the number of classes, are each represented as a \(C\)-dimensional fully connected layer with a softmax activation. All output layers are connected to the second fully connected layer.

As a result, the effect parameter models for the \emph{compressor}, \emph{EQ}, \emph{phaser}, and \emph{hall reverb} effects have a single 3-dimensional continuous output layer and the \emph{distortion} model has two output layers: a 12-dimensional categorical output layer and a 1-dimensional continuous output layer. The architecture of the \emph{distortion} effect parameter model is shown in Figure \ref{fig_distortion_arch}.

\subsection{Effect Selection Model}

The effect selection model outputs a probability distribution of what effect should be applied next to the input audio. As shown in Figure \ref{fig_system}, it takes as input three sequences of the same length corresponding to the number of steps that have been taken by the system. The first two are sequences consisting of effect parameter model inputs: Mel spectrograms and MFCCs. The third is a sequence of 6-dimensional one-hot vectors representing which effect was applied to the input audio at each step. The first five dimensions of the one-hot vector correspond to the five supported effects and the last dimension indicates the initial step of the system when no effect has been applied yet to the input audio. 

Features are extracted from the Mel spectrogram and MFCC sequences using time distributed CNNs identical in architecture to the ones in the effect parameter models, except each convolutional layer uses only half as many filters and there is only one fully connected layer which is the output. The extracted feature vectors are then concatenated with the one-hot vector effect sequence and are then fed through a 128-dimensional bi-directional long short-term memory (LSTM) layer. This is followed by a 128-dimensional fully connected layer with an ELU activation and 50\% dropout applied to it, and lastly a 5-dimensional fully connected output layer with a softmax activation.

\subsection{Training}

One effect parameter model is trained per effect for each preset group (therefore 15 in total) on the Cartesian product of all possible input audio and target audio Mel spectrogram and MFCC pairs. This results in each model being trained on approximately 1.2M data points. Categorical output layers are trained with categorical cross-entropy loss and continuous output layers are trained with mean squared error loss. Multiple losses due to multiple output layers, such as for the \emph{distortion} effect parameter model, are weighted equally and a batch size of 128 is used.

One effect selection model is trained for each preset group (therefore three in total) on unique sequences consisting of one to five steps (applied effects) generated randomly from the available audio clip Mel spectrograms and MFCCs. To encourage efficiency, each effect can only be applied once to the input audio. Consequently, the shortest generated sequence is of length two since this represents the original input audio and one effect being applied and similarly, the longest generated sequence is of length six which represents the original input audio and the five supported effects being applied. Approximately 500k data points are generated for each preset group effect selection model. A categorical cross-entropy loss is used for training along with a batch size of 32.

All models are trained for 100 epochs with early stopping (8 epochs of patience) and a validation and test split of 0.10 and 0.05 respectively. The Adam optimizer \cite{ref_adam} is used with a learning rate of 0.001.

\section{Evaluation}

We evaluate our system at three levels of granularity: the effect parameter models, the effect selection models, and the entire system / ensemble of models as a whole. Each of the three preset groups are evaluated separately and yield very similar and consistent results. We display detailed results for the most challenging preset group, \emph{Advanced Modulating Shapes}, and place the figures and tables for the other two preset groups in the Appendix. Since understanding how perceptually close two audio samples are from metrics alone is difficult, audio examples for \emph{SerumRNN} can be found at \url{https://bit.ly/serum_rnn}.

\subsection{Audio Similarity Metrics}
\label{sect_metrics}

In order to effectively evaluate our system, the distance between two audio clips needs to be computed; specifically the similarity between their timbres. We make use of a variety of six different metrics to evaluate our models vigorously.

The first three metrics are quite basic: the mean squared error (MSE), mean absolute error (MAE), and log-spectral distance (LSD) between two magnitude spectrograms. LSD is defined as the average root mean square difference between each frame of two log power spectrograms.
\begin{equation}
MSE = \frac{1}{mn}\sum_{i=1}^{m}\sum_{j=1}^{n}(S_{ij} - \hat{S}_{ij})^{2}
\end{equation}
\begin{equation}
\label{eq_mae}
MAE = \frac{1}{mn}\sum_{i=1}^{m}\sum_{j=1}^{n}\mid S_{ij} - \hat{S}_{ij} \mid
\end{equation}
\begin{equation}
LSD = \frac{1}{n}\sum_{i=1}^{n}\sqrt{\frac{1}{m}\sum_{j=1}^{m}(S_{ij} - \hat{S}_{ij})^{2}}
\end{equation}
\(S\) is an output decibel Mel spectrogram matrix, \(\hat{S}\) is the target decibel Mel spectrogram matrix, \(m\) is the number of Mel filters, and \(n\) is the number of frames.

The fourth metric (MFCCD) is the mean Euclidean distance between the first 30 MFCCs which represent the smoothed spectral envelope of an audio clip.
\begin{equation}
MFCCD = \frac{1}{n}\sum_{i=1}^{n}\|C_{i} - \hat{C}_{i}\|_{2}
\end{equation}
\(C\) is a matrix of the first 30 output audio MFCCs for each frame, \(\hat{C}\) is a matrix of the first 30 target audio MFCCs for each frame, and \(n\) is the number of frames.

The Pearson Correlation Coefficient (PCC) can also be used to measure audio reconstruction quality \cite{ref_inversynth} and is defined as 
\begin{equation}
PCC = \frac{cov(x,y)}{\sigma_x \sigma_y}
\end{equation}
where \(x\) and \(y\) are 1-dimensional vectors, not matrices. We apply this metric to decibel Mel spectrogram matrices \(S\) and \(\hat{S}\) by flattening each one via row concatenation.

Our last and most robust metric is based off a multi-scale spectral loss \cite{ref_ddsp} that we have modified and we call it multi-scale spectral mean absolute error (MSSMAE). Given an output audio clip and a target audio clip, magnitude spectrograms (\(S_i\) and \(\hat{S_i}\) respectively) are calculated for several different FFT sizes \(i\). We then also compute \(\log S_i\) and \(\log \hat{S_i}\) and define the metric as
\begin{equation}
MSSMAE = \frac{1}{w}\sum_i (MAE(S_i,\hat{S_i}) + \alpha MAE(\log S_i,\log \hat{S_i}))
\end{equation}
where \(w\) is the number of different FFT sizes, \(\alpha\) is a constant scaling term, and MAE is defined in Equation \ref{eq_mae} as the mean squared error between two spectrograms. We set \(\alpha\) to 0.1 in our experiments to make the two MAE values roughly equal since the \(\log\) values are generally around one order of magnitude larger. We use FFT sizes of (64, 128, 256, 512, 1024, 2048) making \(w = 6\) and the neighboring frames in the resulting spectrograms have 75\% overlap. By calculating multiple regular and logarithmic spectrograms with different FFT sizes, two audio clips can be compared along different spatial-temporal resolutions.

For the MSE, MAE, LSD, MFCCD, and MSSMAE metrics, lower values indicate a higher similarity between two audio clips whereas the opposite is true for PCC values. MSE, MAE, PCC, and MSSMAE values are multiplied by 100 for better readability.

\subsection{Effect Parameter Models}

\begin{table}
  \caption{Effect parameter models eval. metrics (\emph{Adv. Mod. Shapes} preset group).}
  \label{table_ams_effect_model_eval}
  \centering
  \begin{tabular}{llrrrr}
    \toprule
    && \multicolumn{4}{c}{Mean Error against Target Audio} \\
    \cmidrule(r){3-6}
    Effect & Metric & Input Audio & Output Audio & $\Delta$ & $\Delta\%$ \\
    \midrule
    \emph{Compressor}  & MSE    &   1.86 &   0.82 &  -1.04 & -56.06 \\
                       & MAE    &  10.50 &   6.43 &  -4.07 & -38.75 \\
                       & LSD    &  10.04 &   6.52 &  -3.53 & -35.10 \\
                       & MFCCD  & 108.53 &  64.48 & -44.05 & -40.59 \\
                       & PCC    &  82.01 &  86.85 &   4.84 &   5.90 \\
                       & MSSMAE &  71.64 &  45.57 & -26.07 & -36.39 \\
    \midrule
    \emph{Distortion}  & MSE    &   4.49 &   0.78 &  -3.71 & -82.66 \\
                       & MAE    &  14.57 &   6.21 &  -8.36 & -57.40 \\
                       & LSD    &  13.90 &   6.41 &  -7.50 & -53.92 \\
                       & MFCCD  & 146.04 &  59.76 & -86.29 & -59.08 \\
                       & PCC    &  74.67 &  80.42 &   5.75 &   7.70 \\
                       & MSSMAE &  94.33 &  65.35 & -28.97 & -30.72 \\
    \midrule
    \emph{Equalizer}   & MSE    &   1.48 &   0.80 &  -0.68 & -45.92 \\
                       & MAE    &   8.76 &   6.29 &  -2.47 & -28.24 \\
                       & LSD    &   8.64 &   6.49 &  -2.14 & -24.82 \\
                       & MFCCD  &  92.52 &  64.17 & -28.35 & -30.64 \\
                       & PCC    &  84.90 &  86.45 &   1.55 &   1.82 \\
                       & MSSMAE &  64.05 &  64.55 &   0.50 &   0.78 \\
    \midrule
    \emph{Phaser}*     & MSE    &   0.86 &   0.77 &  -0.09 & -10.15 \\
                       & MAE    &   6.89 &   6.31 &  -0.58 &  -8.44 \\
                       & LSD    &   7.06 &   6.59 &  -0.47 &  -6.67 \\
                       & MFCCD  &  72.50 &  64.42 &  -8.08 & -11.14 \\
                       & PCC    &  86.05 &  85.58 &  -0.47 &  -0.55 \\
                       & MSSMAE &  51.53 &  47.33 &  -4.20 &  -8.16 \\
    \midrule
    \emph{Hall Reverb} & MSE    &   1.28 &   0.48 &  -0.80 & -62.60 \\
                       & MAE    &   8.32 &   4.89 &  -3.44 & -41.28 \\
                       & LSD    &   8.32 &   5.04 &  -3.28 & -39.43 \\
                       & MFCCD  &  80.49 &  46.21 & -34.28 & -42.59 \\
                       & PCC    &  83.87 &  89.47 &   5.60 &   6.68 \\
                       & MSSMAE &  60.87 &  38.56 & -22.31 & -36.65 \\
    \bottomrule
  \end{tabular}
\end{table}

\begin{table}
  \caption{Effect selection models prediction accuracy (\emph{Adv. Mod. Shapes} preset group).}
  \label{table_ams_next_effect_model_eval}
  \centering
  \begin{tabular}{lrrrrrr}
    \toprule
    & \multicolumn{6}{c}{Step Number (Number of Effects for \emph{SerumCNN})} \\
    \cmidrule(r){2-7}
    Effect Selection Model & 1 & 2 & 3 & 4 & 5 & All \\
    \midrule
    \emph{SerumRNN}    & \textbf{0.994} & \textbf{0.984} & \textbf{0.983} & \textbf{0.979} & 0.999 & \textbf{0.985} \\
    \emph{SerumRNN NI} & 0.992 & 0.974 & 0.961 & 0.962 & \textbf{1.000} & 0.971 \\
    \emph{SerumCNN}    & 0.993 & 0.975 & 0.946 & 0.916 & 0.889 & 0.955 \\
    \bottomrule
  \end{tabular}
\end{table}

Effect parameter models are evaluated using the six metrics defined in Section \ref{sect_metrics} on 1000 input and target audio pairs from their corresponding test data sets. We observe from Table \ref{table_ams_effect_model_eval} that for almost every metric and effect, the models substantially decrease the distance between the input audio clip and the target audio clip. 

*It is worth noting that this trend appears to be less dramatic for the \emph{Phaser} effect, but can be explained by the fact that spectrograms and cepstra do not represent phase information well \cite{ref_ddsp}. As a result, our error metrics might be less representative of the \emph{Phaser} effect when compared to the other effects. 


\subsection{Effect Selection Model}
\label{sect_effect_selection_eval}

We compare the effect selection model, which we call \emph{SerumRNN}, against two baselines to showcase the advantage of 1: using an iterative system and 2: considering the order of effects. 

The first baseline is a non-iterative version of the effect selection model (called \emph{SerumRNN NI}) that uses an identical architecture and training process except that only the first step (the original input and target audio) of the Mel spectrogram and MFCC input sequences is used. As a result, the sequence of effects to be applied to the input audio can be predicted all at once using \emph{SerumRNN NI} without iteratively modifying the audio in between steps using the effect parameter models.

The second baseline is the effect parameter model CNN with a 5-dimensional fully connected output layer and a sigmoid activation. It is trained with a binary cross-entropy loss to predict all effects at once without considering order. We call this baseline \emph{SerumCNN}. 

The accuracy of all three models for effect sequences of length one to five is measured using their test data sets of approximately 25k data points and is summarized in Table \ref{table_ams_next_effect_model_eval}. We observe that all three models perform remarkably well, with \emph{SerumRNN} outperforming the other two for virtually all cases. Since the RNN models predict effects one by one, we believe accuracy is highest for the first and last step due to the first step representing the most dramatic difference between input and target audio clips and the last step being deducible given the previous effects since repeated effects are not supported. The behavior of \emph{SerumCNN} is also expected: accuracy decreases the longer the effect sequence is since more effects need to be predicted correctly.

\subsection{Ensemble / Entire System}

Similarly to the effect parameter models, we evaluate the entire ensemble of models (\emph{SerumRNN}) using the six metrics defined in Section \ref{sect_metrics} on 1000 input and target audio pairs. These pairs are sampled from the unseen test sets and the target audio is the result of a sequence of one to five effects (200 audio pairs per sequence length) being applied to the input audio. Inference time for the entire system (audio processing, effect selection, parameter programming, and audio rendering) is approximately 300 ms per effect applied to the input audio.

\subsubsection{Baselines}
We compare \emph{SerumRNN} against five baselines. The first two are \emph{SerumRNN NI} and \emph{SerumCNN} from Section \ref{sect_effect_selection_eval} which allow us to test for any gains from using an iterative system and considering the order of effects, respectively. The next baseline is a perfect effect selection model (\emph{Oracle}) that always selects the "correct" next effect as is defined by the effect sequence that created the target audio. This allows us to compare our systems to what should be the technically "correct" solution. In order to test the benefit of using an effect selection model in the first place, we also have a \emph{Random} baseline which simply chooses a random effect each time it is called. Finally, to test the benefit of applying the effect parameter models iteratively, we include a baseline called \emph{SerumCNN 1S}. This baseline is identical to the \emph{SerumCNN} baseline except that the effect parameter models are applied all at once to the input audio (after all effects have been predicted at once by \emph{SerumCNN}). This results in a one-shot system with the fastest inference time possible, but no intermediate steps. This baseline is also the most similar to existing neural network synthesizer programming architectures of prior works in related literature \cite{ref_inversynth,ref_ga_2_neural_1}.

All six systems use the same five effect parameter models for each preset group. Since repeated effects are not supported, if an effect selection model chooses an effect that has already been applied in a previous step, the next most probable unused effect is selected. As a result, all systems terminate at the latest after applying the maximum of five supported effects.

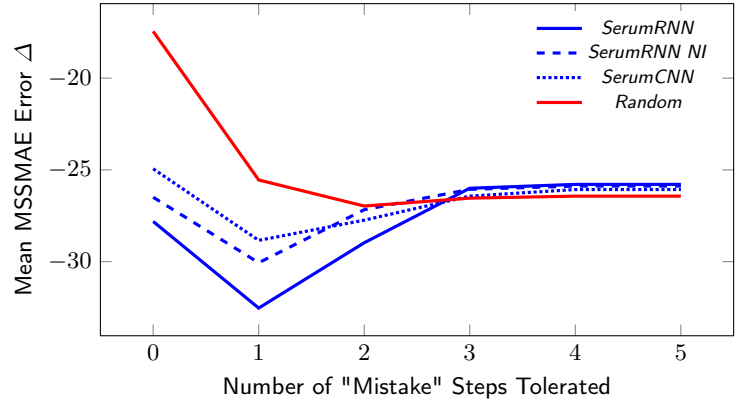
\begin{figure}[h]
\centering
\begin{tikzpicture}[font=\sffamily]
\begin{axis}[
	xlabel=Number of "Mistake" Steps Tolerated,
	ylabel=Mean MSSMAE Error $\Delta$,
	width=10cm,height=6cm,
    legend style={draw=none, nodes={scale=0.8, transform shape}}
]


\addplot[solid, line width=1.2pt, color=blue] coordinates {
	(0, -27.81)
	(1, -32.53)
	(2, -28.97)
	(3, -26.00)
	(4, -25.79)
	(5, -25.79)
};

\addplot[dashed, line width=1.2pt, color=blue] coordinates {
	(0, -26.49)
	(1, -30.06)
	(2, -27.17)
	(3, -26.06)
	(4, -25.88)
	(5, -25.88)
};

\addplot[densely dotted, line width=1.2pt, color=blue] coordinates {
	(0, -24.94)
	(1, -28.84)
	(2, -27.74)
	(3, -26.43)
	(4, -26.07)
	(5, -26.07)
};

\addplot[line width=1.2pt, color=red] coordinates {
	(0, -17.45)
	(1, -25.55)
	(2, -26.97)
	(3, -26.54)
	(4, -26.43)
	(5, -26.43)
};


\legend{\emph{SerumRNN}, \emph{SerumRNN NI}, \emph{SerumCNN}, \emph{Random}}

\end{axis}
\end{tikzpicture}
\caption{Relationship between the tolerated number of mistake steps vs. the overall\\ error $\Delta$ against the target audio (\emph{Adv. Mod. Shapes} preset group).}
\label{fig_ams_until_worse_graph}
\end{figure}

\subsubsection{Stopping Condition}

Before the systems can be evaluated, a stopping condition must be defined for \emph{SerumRNN}, \emph{SerumRNN NI}, \emph{SerumCNN}, and \emph{Random}. Otherwise five effects will always be applied to the input audio, even when the target audio was created using only one effect. We believe a reasonable stopping condition is terminating the system once it makes a mistake, i.e. when the distance between the input and target audio increases. However, there may be situations where making a "mistake" is advantageous since the ideal sequence of steps to the target audio can be non-monotonic. Therefore, we investigate in Figure \ref{fig_ams_until_worse_graph} how the number of tolerated bad steps affects the resulting overall error reduction of each system. From the plotted results we can conclude that the ideal stopping condition is tolerating one step in the wrong direction before terminating.

\subsubsection{Discussion}

Table \ref{tab_ams_all_eval} summarizes the evaluation results of \emph{SerumRNN} compared to the five baselines and Table \ref{table_ams_serum_rnn_eval} gives more granular evaluation details for \emph{SerumRNN} and the \emph{Advanced Modulating Shapes} preset group.

All systems significantly reduce the error between the input and target audio which indicates that the effect parameter models individually are robust and highly effective, even when applied in a random order. We believe using two channel Mel spectrogram and MFCC inputs and training on the Cartesian product of possible input and target audio combinations (as explained in Section \ref{sect_effect_param_modeling}) helped achieve this robustness.

Perhaps unsurprisingly, \emph{SerumRNN} consistently outperforms \emph{SerumRNN NI}, \emph{SerumCNN}, and \emph{SerumCNN 1S} for all metrics. This highlights the importance of considering the order of effects and using an iterative approach for both effect selection and effect parameter programming. Performance of the one-shot \emph{SerumCNN 1S} system is also quite good considering it is up to five times faster than any of the other systems. If inference speed is critically important, a reasonable compromise can be made by choosing to use this system instead.

We also note for \emph{SerumRNN} that all five steps reduce the error between the input and target audio clips, with the largest gains achieved in the first two steps and the smallest in the final two steps. In contrast to this, for the \emph{Random} baseline (Table \ref{table_ams_random_eval}) error reduction is distributed essentially evenly between the five steps. This indicates that the effect selection model chooses the effects in order of importance for multi-effect sequences, thus making the intermediate steps of \emph{SerumRNN} more educational for the user.

The most surprising result is that \emph{SerumRNN} consistently outperforms the Oracle baseline, especially for the most robust MSSMAE metric. This implies that \emph{SerumRNN} sometimes chooses effects in an order that is more optimal for the effect parameter models than using the original effect order that created the target audio. We find this result exciting because it suggests our neural architecture and ensemble of models is able to extrapolate and discover solutions outside of the scope of what would be considered "correct" by the training data.

From an auditory point of view, the reduction in error for each step of \emph{SerumRNN} can be heard very clearly with the final output audio often being indistinguishable from the target audio. We highly recommend listening to the audio examples at \url{https://bit.ly/serum_rnn}.

\begin{table}[h]
  \caption{Comparison of overall mean error reduction for all systems.}
  \label{tab_ams_all_eval}
  \centering
  \begin{tabular}{llrrrrrr}
    \toprule
    & & \multicolumn{6}{c}{Mean Error $\Delta$ against Target Audio} \\
    \cmidrule(r){3-8}
    Preset Group & System & MSE & MAE & LSD & MFCCD & PCC & MSSMAE \\
    \midrule
    \emph{Advanced} & Oracle & -4.02 & \textbf{-9.45} & \textbf{-8.48} & \textbf{-97.14} & 14.80 & -28.67 \\
    \emph{Modulating} & \emph{SerumRNN} & \textbf{-4.04} & -9.41 & -8.36 & -96.17 & \textbf{15.53} & \textbf{-32.53} \\
    \emph{Shapes} & \emph{SerumRNN NI}  & -4.02 & -9.20 & -8.25 & -94.55 & 15.26 & -30.06 \\
    & \emph{SerumCNN}          & -3.97 & -9.21 & -8.18 & -93.72 & 14.71 & -28.84 \\
    & Random                   & -3.33 & -7.45 & -6.52 & -74.40 & 11.96 & -25.55 \\
    & \emph{SerumCNN 1S} & -3.96 & -9.12 & -8.10 & -92.68 & 14.51 & -28.50 \\
    \midrule
    \emph{Advanced} & Oracle        & -2.96 & \textbf{-8.09} & \textbf{-7.21} & -83.01 & 15.90 & -37.74 \\
    \emph{Shapes} & \emph{SerumRNN} & \textbf{-2.97} & -8.08 & -7.17 & \textbf{-83.34} & \textbf{16.87} & \textbf{-41.24} \\
    & \emph{SerumRNN NI}            & -2.96 & -7.98 & -7.07 & -81.80 & 16.79 & -39.14 \\
    & \emph{SerumCNN}               & -2.92 & -7.84 & -7.05 & -81.21 & 15.84 & -38.89 \\
    & Random                        & -2.48 & -6.39 & -5.71 & -65.95 & 14.20 & -35.18 \\
    & \emph{SerumCNN 1S}            & -2.88 & -7.65 & -6.83 & -78.40 & 15.55 & -37.80 \\
    \midrule
    \emph{Basic Shapes} & Oracle    & -4.00 & -9.49 & -8.19 & -92.12 & 18.78 & -41.16 \\
    & \emph{SerumRNN}               & \textbf{-4.08} & \textbf{-9.62} & \textbf{-8.35} & \textbf{-94.08} & \textbf{20.03} & \textbf{-42.32} \\
    & \emph{SerumRNN NI}            & -3.79 & -9.02 & -7.75 & -86.36 & 19.23 & -40.67 \\
    & \emph{SerumCNN}               & -3.45 & -8.50 & -7.25 & -81.01 & 19.32 & -41.95 \\
    & Random                        & -3.14 & -7.20 & -6.21 & -69.70 & 15.14 & -35.30 \\
    & \emph{SerumCNN 1S}            & -3.56 & -8.52 & -7.36 & -82.60 & 18.93 & -40.06 \\
    \bottomrule
  \end{tabular}
\end{table}

\begin{table}[h!]
  \caption{Eval. metrics for \emph{SerumRNN} (\emph{Adv. Mod. Shapes} preset group).}
  \label{table_ams_serum_rnn_eval}
  \centering
  \begin{tabular}{l}
    \toprule
    \multicolumn{1}{c}{} \\
    \cmidrule(r){1-1}
    Metric \\
    \midrule
    MSE \\
    MAE \\
    LSD \\
    MFCCD \\
    PCC \\
    MSSMAE \\
    \bottomrule
  \end{tabular}
  \begin{tabular}{rrrr}
    \toprule
    \multicolumn{4}{c}{Mean Error against Target} \\
    \cmidrule(r){1-4}
    Init. & Final & $\Delta$ & $\Delta\%$ \\
    \midrule
      4.97 &  0.94 &  -4.04 & -81.19 \\
     16.47 &  7.06 &  -9.41 & -57.11 \\
     15.59 &  7.23 &  -8.36 & -53.63 \\
    167.52 & 71.35 & -96.17 & -57.41 \\
     68.48 & 84.01 &  15.53 &  22.68  \\
     83.38 & 50.85 & -32.53 & -39.02 \\
    \bottomrule
  \end{tabular}
  \begin{tabular}{rrrrr}
    \toprule
    \multicolumn{5}{c}{Mean Error $\Delta$ per Step} \\
    \cmidrule(r){1-5}
    1 & 2 & 3 & 4 & 5 \\
    \midrule
     -2.64 &  -0.86 &  -0.38 &  -0.16 &  -0.10 \\
     -5.49 &  -2.30 &  -1.13 &  -0.48 &  -0.33 \\
     -5.04 &  -1.95 &  -0.97 &  -0.41 &  -0.27 \\
    -56.37 & -23.26 & -11.41 &  -5.12 &  -3.19 \\
      5.81 &   6.23 &   2.37 &   1.33 &   1.53 \\
     -3.70 & -15.63 &  -9.47 &  -3.81 &  -2.23 \\
    \bottomrule
  \end{tabular}
\end{table}

\begin{table}[h]
  \caption{Evaluation metrics for the Random baseline (\emph{Adv. Mod. Shapes} preset group).}
  \label{table_ams_random_eval}
  \centering
  \begin{tabular}{l}
    \toprule
    \multicolumn{1}{c}{} \\
    \cmidrule(r){1-1}
    Metric \\
    \midrule
    MSE \\
    MAE \\
    LSD \\
    MFCCD \\
    PCC \\
    MSSMAE \\
    \bottomrule
  \end{tabular}
  \begin{tabular}{rrrr}
    \toprule
    \multicolumn{4}{c}{Mean Error against Target} \\
    \cmidrule(r){1-4}
    Init. & Final & $\Delta$ & $\Delta\%$ \\
    \midrule
      4.99 &   1.66 &  -3.33 & -66.71 \\
     16.49 &   9.05 &  -7.45 & -45.15 \\
     15.62 &   9.10 &  -6.52 & -41.72 \\
    167.81 &  93.41 & -74.40 & -44.34 \\
     68.24 &  80.20 &  11.96 &  17.52 \\
     83.16 &  57.60 & -25.55 & -30.73 \\
    \bottomrule
  \end{tabular}
  \begin{tabular}{rrrrr}
    \toprule
    \multicolumn{5}{c}{Mean Error $\Delta$ per Step} \\
    \cmidrule(r){1-5}
    1 & 2 & 3 & 4 & 5 \\
    \midrule
     -0.84 &  -0.95 &  -0.85 &  -0.81 &  -0.69 \\
     -1.70 &  -2.00 &  -1.98 &  -1.99 &  -1.89 \\
     -1.47 &  -1.74 &  -1.78 &  -1.70 &  -1.69 \\
    -16.85 & -19.98 & -19.84 & -19.26 & -19.53 \\
      1.73 &   3.72 &   3.85 &   4.00 &   4.15 \\
     -4.04 &  -6.34 &  -7.63 &  -5.91 &  -8.74 \\
    \bottomrule
  \end{tabular}
\end{table}

\section{Conclusion}

Overall, \emph{SerumRNN} is consistently able to produce intermediate steps that bring the input audio significantly closer to the target audio while using a fully featured, industry leading synthesizer VST plugin. \emph{SerumRNN} also provides near real-time, quantitative feedback about which effects are the most important. In addition to this, it can also find effect sequence orders that are potentially more optimal than what the target audio was originally created with. With \emph{SerumRNN} the user can pick and choose which intermediate steps they would like to use and can feed tweaked versions back into the system for additional fine-tuning or to learn more. We also noticed fun creative applications when our system produced unexpected results or was given significantly out of domain target audio.

Currently, the training data for \emph{SerumRNN} is sampled randomly and thus many of the generated sounds are not useful musically. As future work we would like to improve the efficiency of this using adversarial techniques. We would also like to evaluate the system's usefulness and perceived audio similarity with a user study and plan on packaging \emph{SerumRNN} into a Max for Live \cite{ref_max_for_live} plugin to make it easily accessible to anyone learning sound design.

We believe our research is just the tip of the iceberg for building AI powered sound design tools. We can imagine a future where tools like ours might be able to find more efficient and simpler methods of creating sounds, thus educating students more effectively and democratizing sound design.

%
%
%
\bibliographystyle{splncs04}
\bibliography{mybibliography}

\appendix
\section{Appendix}

\begin{table}[h]
  \caption{Effect selection models prediction accuracy (\emph{Advanced Shapes} and\\ \emph{Basic Shapes} preset groups).}
  \label{table_as_next_effect_model_eval}
  \centering
  \begin{tabular}{llrrrrrr}
    \toprule
    & & \multicolumn{6}{c}{Step Number (No. of Effects for \emph{SerumCNN})} \\
    \cmidrule(r){3-8}
    Preset Group & Effect Sel. Model & 1 & 2 & 3 & 4 & 5 & All \\
    \midrule
    \emph{Advanced Shapes} & \emph{SerumRNN}    & \textbf{0.993} & \textbf{0.983} & \textbf{0.981} & \textbf{0.986} & 0.994 & \textbf{0.985} \\
    & \emph{SerumRNN NI} & 0.990 & 0.968 & 0.959 & 0.966 & \textbf{1.000} & 0.969 \\
    & \emph{SerumCNN}    & 0.992 & 0.968 & 0.941 & 0.921 & 0.909 & 0.952 \\
    \midrule
    \emph{Basic Shapes} & \emph{SerumRNN}    & \textbf{0.992} & \textbf{0.983} & \textbf{0.983} & \textbf{0.974} & \textbf{1.000} & \textbf{0.983} \\
    & \emph{SerumRNN NI} & 0.991 & 0.970 & 0.954 & 0.971 & \textbf{1.000} & 0.969 \\
    & \emph{SerumCNN}    & 0.990 & 0.963 & 0.936 & 0.913 & 0.884 & 0.947 \\
    \bottomrule
  \end{tabular}
\end{table}


\begin{table}[h]
  \caption{Evaluation metrics for \emph{SerumRNN} (\emph{Advanced Shapes} preset group).}
  \label{table_as_serum_rnn_eval}
  \centering
  \begin{tabular}{l}
    \toprule
    \multicolumn{1}{c}{} \\
    \cmidrule(r){1-1}
    Metric \\
    \midrule
    MSE \\
    MAE \\
    LSD \\
    MFCCD \\
    PCC \\
    MSSMAE \\
    \bottomrule
  \end{tabular}
  \begin{tabular}{rrrr}
    \toprule
    \multicolumn{4}{c}{Mean Error against Target} \\
    \cmidrule(r){1-4}
    Init. & Final & $\Delta$ & $\Delta\%$ \\
    \midrule
      3.92 &   0.95 &  -2.97 & -75.80 \\
     15.28 &   7.20 &  -8.08 & -52.87 \\
     14.44 &   7.27 &  -7.17 & -49.64 \\
    153.24 &  69.90 & -83.34 & -54.38 \\
     64.26 &  81.12 &  16.87 &  26.25  \\
     95.76 &  54.52 & -41.24 & -43.07 \\
    \bottomrule
  \end{tabular}
  \begin{tabular}{rrrrr}
    \toprule
    \multicolumn{5}{c}{Mean Error $\Delta$ per Step} \\
    \cmidrule(r){1-5}
    1 & 2 & 3 & 4 & 5 \\
    \midrule
     -1.74 &  -0.76 &  -0.35 &  -0.14 &  -0.06 \\
     -4.48 &  -2.01 &  -1.13 &  -0.51 &  -0.20 \\
     -4.12 &  -1.68 &  -0.96 &  -0.45 &  -0.17 \\
    -46.05 & -20.51 & -11.57 &  -5.89 &  -2.34 \\
      7.59 &   5.63 &   2.32 &   1.75 &   1.48 \\
    -19.55 & -11.00 &  -5.82 &  -4.16 &  -2.29 \\
    \bottomrule
  \end{tabular}
\end{table}

\begin{table}[h!]
  \caption{Evaluation metrics for \emph{SerumRNN} (\emph{Basic Shapes} preset group).}
  \label{table_bs_serum_rnn_eval}
  \centering
  \begin{tabular}{l}
    \toprule
    \multicolumn{1}{c}{} \\
    \cmidrule(r){1-1}
    Metric \\
    \midrule
    MSE \\
    MAE \\
    LSD \\
    MFCCD \\
    PCC \\
    MSSMAE \\
    \bottomrule
  \end{tabular}
  \begin{tabular}{rrrr}
    \toprule
    \multicolumn{4}{c}{Mean Error against Target} \\
    \cmidrule(r){1-4}
    Init. & Final & $\Delta$ & $\Delta\%$ \\
    \midrule
      5.25 &   1.16 &  -4.08 & -77.83 \\
     16.82 &   7.20 &  -9.62 & -57.19 \\
     15.69 &   7.34 &  -8.35 & -53.25 \\
    166.76 &  72.68 & -94.08 & -56.42 \\
     65.97 &  86.00 &  20.03 &  30.37  \\
     87.18 &  44.85 & -42.32 & -48.55 \\
    \bottomrule
  \end{tabular}
  \begin{tabular}{rrrrr}
    \toprule
    \multicolumn{5}{c}{Mean Error $\Delta$ per Step} \\
    \cmidrule(r){1-5}
    1 & 2 & 3 & 4 & 5 \\
    \midrule
     -2.58 &  -0.93 &  -0.39 &  -0.21 &  -0.08 \\
     -5.80 &  -2.23 &  -1.05 &  -0.66 &  -0.16 \\
     -5.09 &  -1.86 &  -0.92 &  -0.57 &  -0.18 \\
    -55.78 & -22.05 & -10.78 &  -6.57 &  -2.03 \\
     13.02 &   3.33 &   2.06 &   1.76 &   1.19 \\
    -21.67 & -13.33 &  -3.46 &  -3.36 &  -1.95 \\
    \bottomrule
  \end{tabular}
\end{table}

\end{document}